\documentclass[%
 reprint,
 superscriptaddress,
 amsmath,amssymb,
 aps,
 prb,
floatfix,
]{revtex4-1}

\usepackage{natbib}
\usepackage{float}
\usepackage{graphicx}
\usepackage{color}
\usepackage[colorlinks=true,linkcolor=blue]{hyperref}
\usepackage{dcolumn}
\usepackage{bm}

\begin{document}

\title{Multiple energy-scales in vertex-frustrated mesospin systems}

\author{Henry Stopfel}%
\email[Email:]{henry.stopfel@physics.uu.se}
\affiliation{%
 Department of Physics and Astronomy, Uppsala University, Box 516, SE-75120, Uppsala, Sweden
}%
\author{Unnar~B.~Arnalds}%
\affiliation{%
 Science Institute, University of Iceland, Reykjavik, Iceland
}%
\author{Aaron Stein}%
\affiliation{%
 Center for Functional Nanomaterials, Brookhaven National Laboratory, Upton, New York 11973, USA
}%
\author{Thomas~P.~A.~Hase}
\affiliation{
 Department of Physics, University of Warwick, Coventry, United Kingdom
}%
\author{Bj\"{o}rgvin Hj\"{o}rvarsson}%
\affiliation{%
 Department of Physics and Astronomy, Uppsala University, Box 516, SE-75120, Uppsala, Sweden
}%
\author{Vassilios Kapaklis}%
\email[Email:]{vassilios.kapaklis@physics.uu.se}
\affiliation{%
 Department of Physics and Astronomy, Uppsala University, Box 516, SE-75120, Uppsala, Sweden
}%

\date{\today}

\begin{abstract}
    The interplay between topology and energy-hierarchy plays a vital role in the collective magnetic order in artificial ferroic systems. Here we investigate, experimentally, the effect of having one or two activation energies of interacting Ising-like magnetic islands -- mesospins -- in thermalized, vertex-frustrated lattices. The thermally arrested magnetic states of the elements were determined using synchrotron-based magnetic microscopy after cooling the samples from temperatures above the Curie temperature of the material.  Statistical analysis of the correlations between mesospins across several length-scales, reveals changes in the magnetic order, reflecting the amount of ground state plaquettes realized for a vertex-frustrated lattice. We show that the latter depends on the presence, or not, of different activation energies.
\end{abstract}

\maketitle

\section{Introduction}
 
Tailoring the field response of magnetic mesospins using a hierarchy of energy scales, was originally demonstrated by \citet{Cowburn2000Sci}. In their case, the hierarchy was obtained from a distribution in the size and shape of small magnetic elements which facilitated the design of magnetic cellular automata, with potential uses in information processing and non-volatile information storage \cite{Cowburn2000Sci, Cowburn_IOP_1999, Imre_Science_2006}. Even though these structures were thermally inactive, their results highlighted the importance of the hierarchy of energies on the observed magnetic order and global response of the system. Since then, the exploration of the magnetic properties of nano-arrays has expanded dramatically \cite{HeydermanStamps_review_2013,Nisoli_Kapaklis_NatPhys2017,Rougemaille_review_2019} and now includes arrays designed to exhibit geometrical \cite{Wang2006Nat,Canals_2016_NatComm,Perrin2016Nat,Ostman_NatPhys_2018,Farhan_2019_SciAdv} and vertex-frustration \cite{Chern2013PRL, Morrison2013NJP, Gilbert_Shakti_NatPhys2014, Gilbert_tetris_2015}.
The pairwise frustration of the interactions between spins at the vertex level is preserved in the kagome geometry \cite{Canals_2016_NatComm, Nisoli_Kapaklis_NatPhys2017}. In lattice architectures based on square artificial spin ice, however, the pairwise frustration is lifted by the different interaction strengths between parallel and perpendicular islands at the vertex level \cite{Wang2006Nat,Moller2006PRL,Thonig2014}. In recent years, the pairwise frustration for square lattices has been successfully restored by a variety of approaches \cite{Perrin2016Nat,Ostman_NatPhys_2018,Farhan_2019_SciAdv}. A different approach was proposed by Morrison et al. \cite{Morrison2013NJP}, creating lattices where it is impossible to choose all vertices to be in their lowest energy configuration, due to topological constraints. As a result, these vertex-frustrated structures always contain excited vertices, leading to residual entropy. In addition to tailoring the lattice and geometry of the elements, the materials from which the arrays are fabricated can be chosen to allow thermal fluctuations at suitable temperatures \cite{Sloetjes_arXiv_2020}. Such approaches have already enabled the investigation of spontaneous ordering, dynamics and phase transitions on the mesoscale  \cite{Morgan2010NatPhys, Kapaklis2012NJP, Arnalds2012APL, Farhan2013NatPhys, Kapaklis2014NatNano, Anghinolfi_2015_NatComm, SciRep_Relaxation_2016, Ostman_NatPhys_2018, ShaktiME, Sendetskyi_2019_PRB, Pohlit_2020_Susceptibility_PRB, Leo_2020arXiv, Mellado_arXiv_2020}.

Currently there are only a couple of lattices that have been investigated, that incorporate multiple element sizes \cite{Chern2013PRL,Gilbert_Shakti_NatPhys2014,Ostman_NatPhys_2018, ShaktiME}. In \citet{Ostman_NatPhys_2018}, small circular elements were placed between the islands within the vertex and used as interaction modifiers to modify the overall energy landscape of the lattice, but their actual magnetic state/orientation was not determined. 
Even though a difference in activation energy for these two type of mesospins is to be expected, it is not trivial to disentangle the actual contribution of the former, since the  dimensionality between the two mesospin types also differs (Ising- and XY-like) \cite{BjornPRB}. To overcome this shortcoming, systems where activation energies may differ whilst maintaining the same mesospin dimensionality, can serve as ideal settings for the detailed study of the multiple energy-scale impact on the derived magnetic order. The first investigations of this kind were performed on the Shakti lattice \cite{ShaktiME}, which until then had not been addressed with respect to the differently sized elements within the lattice \cite{Chern2013PRL,Gilbert_Shakti_NatPhys2014}. However, the highly degenerate ground state, together with the high symmetry of the Shakti lattice, masks possible longer-range correlations and ordering, limiting the extent of these earlier studies. Thus, other lattice geometries are needed to probe the effect of energy hierarchy on the ordering within artificial spin ice structures, especially on length-scales beyond the nearest neighbour.

\begin{figure*}[t!]
\centering
\includegraphics[width=2\columnwidth]{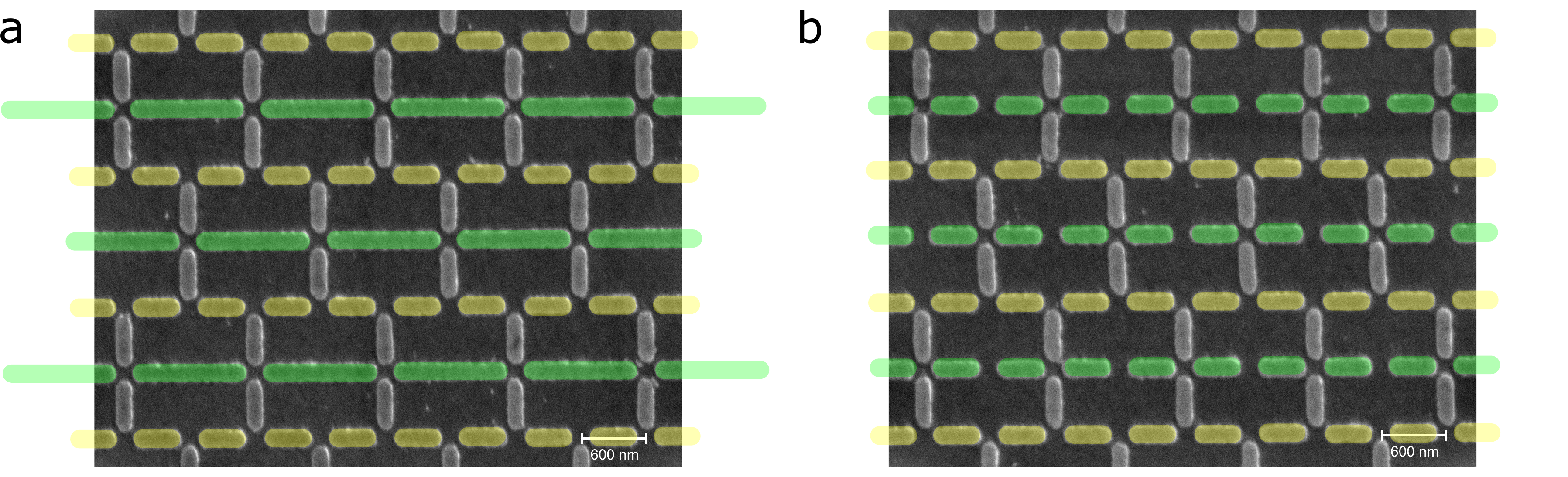}
\caption{{\bf Illustration of the Saint George lattices}. {\bf a} The Saint George (SG) lattice with its two different island ({\it mesospin}) sizes. {\bf b} The modified Saint George (mSG) lattice has only one island size. In the Saint George lattice geometry there are two different horizontal chains, coloured green and yellow. These horizontal chains are connected by vertical islands. The green island chains are referred to as long-mesospin chains in the main text, as they are consisting of the long-islands in the SG lattice and have double the amount of islands in the mSG. The yellow island chains are equal in the SG and mSG lattice.}
\label{fig:SG}
\end{figure*}

In this work, we address the effect on the magnetic order of having one or two activation energies influencing the magnetic correlations among Ising like mesospins.  We do this experimentally, using mesospins within the Saint George (SG) and modified Saint George (mSG) structures, defined and illustrated in Fig.~\ref{fig:SG}. The SG and mSG structures are closely related to Shakti lattices described by \citet{Chern2013PRL}, and constructed in a similar fashion as introduced by \citet{Morrison2013NJP}: removing and merging elements, whilst using the square artificial spin ice lattice as a starting base. As a result, these lattices become vertex-frustrated \cite{Morrison2013NJP}, the SG lattice geometry is characterised by its horizontal Ising-like mesospin chains \cite{Ising1925ZfP, Ostman_Ising_JPCM_2018}. There are two horizontal chain types: one composed of long-mesospins and the other with short-mesospins. Both chain types are coupled via short vertical mesospins which creates an axial anisotropy in the lattice, as can be seen in Fig.~\ref{fig:SG}. The SG lattice has two activation energies, one for the long- and another for the short-mesospins. All islands forming the mSG lattice have the same size and therefore the elements have one and the same intrinsic activation energy.

The ratio of the activation energies within the SG lattice design can give rise to two distinct scenarios: (1)	The two energy-scales are well separated. This results in a freezing of the long-mesospins, while  the short-mesospins remain thermally active. Lowering the temperature further results in a freezing of the short-mesospins, influenced by a static magnetic field originating from the frozen long-islands. (2) The energy-scales of the short- and long-mesospins are close or overlapping. This results in an interplay between the differently sized elements during the freezing process. Both scenarios would give rise to an emergent magnetic order, but the correlations between the magnetic mesospins should be different for the two cases. When the activation energies are very different, one may naively expect the long-mesospins to behave as an independent subset array,  which due to their shape anisotropy and close separation, act as an Ising chain.

\section{Methods}

\subsection{Sample preparation}
 
The magnetic nano-structures were fabricated by a post-growth patterning process on a thin film of $\delta$-doped Palladium (Iron) \cite{Parnaste2007}. $\delta$-doped Palladium (Iron) is in fact a tri-layer of Palladium (Pd) -- Iron (Fe) -- Palladium (Pd), where the Curie-temperature, $T_{\rm{C}}$ and the thermally active temperature range for mesospins is defined by the thickness of the Iron layer and their size \cite{ShaktiME,Arnalds2014APL}. In the present study the nominal Fe thickness is 2.0 monolayers, characterised by a  Curie-temperature of $T_{\rm{C}}=400$~K \cite{Parnaste2007,Papaioannou2010JPCM} for the continuous films. The post-patterning was carried out at the Center for Functional Nanomaterials (CFN), Brookhaven National Laboratory in Upton, New York. All investigated structures for the present study as well as the investigations reported in \citet{ShaktiME}, were fabricated on the same substrate from the same $\delta$-doped Pd(Fe) film, ensuring identical intrinsic material properties as well as the same thermal history for all the investigated structures while performing magnetic imaging. The lengths of the stadium shaped short- and long-mesospins were 450 nm and 1050 nm respectively, while
the width was kept at 150 nm for both. The lattice spacing between two
parallel neighbouring short-elements was chosen to be 600 nm. For more details on the patterning process,  see \citet{ShaktiME}.


\begin{figure*}[t!]
\centering
\includegraphics[width=1.9\columnwidth]{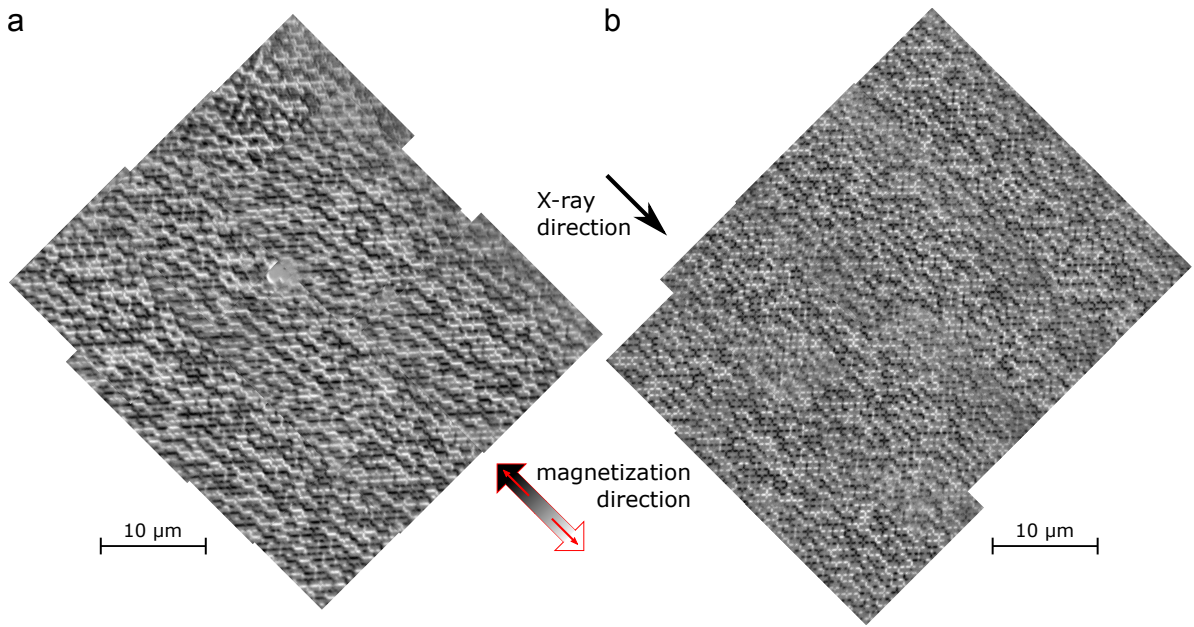}
\caption{{\bf Photoemission electron microscopy images using the x-ray magnetic circular dicroism effect to visualize the magnetization direction.} Combined measurements of {\bf a} the SG lattice and {\bf b} of the mSG lattice. The magnetization direction of each individual island can be determined by its colour. Black illustrated mesospins pointing left or up while white mesospins have their magnetization pointing right or down.}
\label{fig:PEEM}
\end{figure*}

\subsection{Thermal protocol}
 
The thermal protocol involves gradual cooling from a superparamagnetic state of the patterned elements, to an arrested state at the lowest temperatures. Similar to previous studies on the Shakti lattice \cite{ShaktiME} and unlike the investigations on square artificial spin ice \cite{Kapaklis2012NJP,Kapaklis2014NatNano} the SG lattice exhibits two distinctive thermally active regimes, caused by the different sizes of the elements. Below the Curie temperature, ($T_{\rm{c}}$),  the elements are magnetic and considered as mesospins. The temperature-dependent magnetostatic coupling influences the activation of the elements and is therefore biasing one of the two magnetization directions, depending on the states of the adjacent mesospins.

\subsection{Determination of the magnetization direction}
 
The magnetic state of each mesospin was determined by Photo Emission Electron Microscopy (PEEM) using the X-ray Magnetic Circular Dichroism (XMCD) contrast. The experimental studies were performed at the 11.0.1 PEEM3 beamline of the Advanced Light Source, in Berkeley, California, USA. The islands were oriented 45 degrees with respect to the incoming X-ray beam. Multiple PEEM-XMCD images were acquired and stitched together to form an extended image of more than 5000 mesospins. The PEEM-XMCD images were taken at 65~K, far below the Curie-temperature and the blocking temperatures of the mesospins. The PEEM-XMCD images were obtained with a sampling time of $t_{s} = 360~s$, which defines the time  resolution of the measurements: At temperatures far below the blocking temperature of the elements, all magnetic states are stable (frozen) and can be imaged for long periods of time. On the other extreme (higher temperatures), when the magnetization reversal times of the mesospins are much smaller than $t_{s}$ and the elements change their magnetic orientation during the duration of the measurement, no magnetic contrast is obtained.

\subsection{Activation energies}

The intrinsic activation energies of the mesospins are known from earlier investigations, see \citet{ShaktiME}: $E_{\mathrm{short}}/k_B = 560(40)$ K and $E_{\mathrm{long}}/k_B = 1140(140)$ K. The ratio of the activation energies of $non-interacting$ short- and long-mesospins is therefore 0.49(4). The Shakti lattices from \citet{ShaktiME} and the ones discussed here were fabricated on the same chip and measured during the same beamtime at PEEM3, removing any uncertainty concerning the intrinsic properties of the material. The ratio of the activation energies of the short- and the long-mesospins represents a substantial difference and therefore motivates our investigation of the influence of the energy-hierarchy on the magnetic ordering mechanism in such multiple energy-scale systems.

\section{Results and discussions}

\begin{figure}[b!]
\centering
\includegraphics[width=\columnwidth]{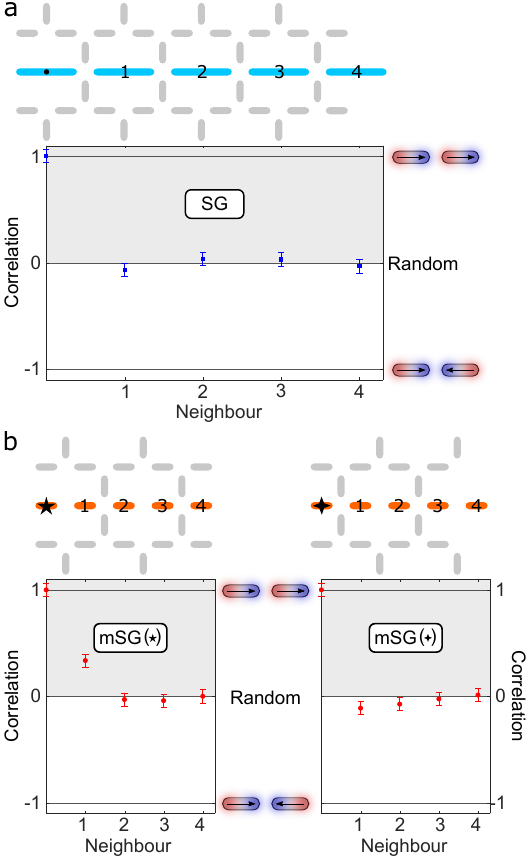}
\caption{{\bf Magnetic order of long-mesospin chains in the SG and mSG lattice. a} Correlation between neighbouring long-mesospins in the SG lattice shows an almost random alignment for the first mesospin neighbour, the slight negative correlation value indicates a  preference towards anti-ferromagnetic order. {\bf b} The same mesospin chains are also investigated in the mSG, where two short-mesospins resemble one long-mesospin. Depending on the starting element ($\star$~or~$+$), the first mesospin neighbour has a positive value ($\star$) and therefore shows a preference for ferromagnetic alignement, or a negative value ($+$) and therefore a slight preference for anti-ferromagnetic alignment.}
\label{fig:SGLongChains}
\end{figure}

The results from photoemission electron microscopy (PEEM-XMCD) experiments for both the SG and mSG lattice are illustrated in Fig.~\ref{fig:PEEM}. While the magnetization direction of each mesospin can be determined, it is difficult to  identify any correlations in and/or between the SG and mSG lattices from a visual inspection of Fig.~\ref{fig:PEEM}. 
To explore the possible effect of different energy hierarchies requires, therefore, a statistical analysis, as described here below.

\subsection{Long-mesospin chains}

In Fig.~\ref{fig:SGLongChains} we show the correlation between the magnetisation direction of the neighbouring long islands,  after cooling the sample from room temperature to 65 K. The correlation in the magnetization direction of neighbouring islands is calculated using: 

\begin{equation}
\label{eq:correlation}
    \centering
    C^n = \dfrac{\sum_k^N m_k \cdot m_{n_k}}{N}.
\end{equation}

\noindent
Here $N$ is the total number of long-mesospins, $m_k$ the magnetization direction of the reference mesospin and $m_{n_k}$ the magnetization direction of the $n^{th}$ neighbour of mesospin $k$.
In this formalism, $m_k$ and $m_{n_k}$ can be $\pm 1$ depending on their magnetization direction (black or white in Fig.~\ref{fig:PEEM}). To determine the correlations in the mSG lattice, the same approach using Equation~\eqref{eq:correlation} was applied, but due to the different local starting condition (identified in Fig.~\ref{fig:SGLongChains}b with $\star$~or~$+$) the correlations were determined separately for each condition. For the long-mesospins, the correlation of the first neighbour is close to be zero, indicating a random arrangement of the long-mesospins. This is in contrast to the ferromagnetic order which would have been naively expected \cite{Ostman_Ising_JPCM_2018}, calling for an analysis of the vertex states.

\begin{figure}[b!]
\centering
\includegraphics[width=\columnwidth]{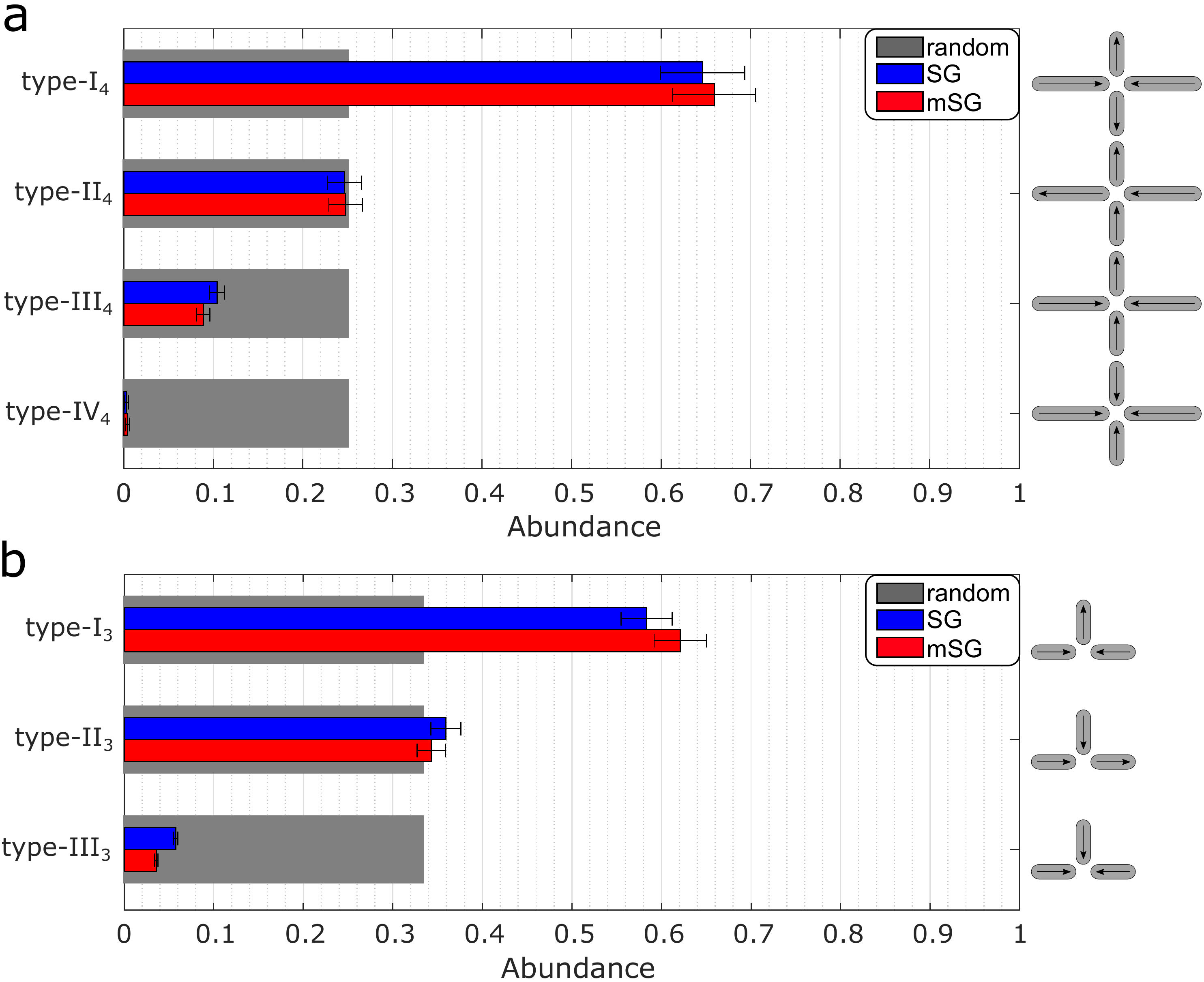}
\caption{{\bf Vertex statistics for the SG and mSG lattice}. {\bf a} The degeneracy corrected vertex abundance for the fourfold coordinated vertices in the SG (blue) and mSG (red) lattice. {\bf b} The degeneracy corrected vertex abundance for the threefold coordinated vertices in the SG (blue) and mSG (red) lattice. The gray bars illustrate the vertex abundance for a random distribution as it would be received if the mesospins would not interact. The vertex type energies are corresponding to their notation, namely type-I$_{4(3)}$ being the lowest energy and therefore the ground state for the fourfold (threefold) coordinated vertex.}
\label{fig:Stats}
\end{figure}

\begin{figure*}[t!]
\centering
\includegraphics[width=2\columnwidth]{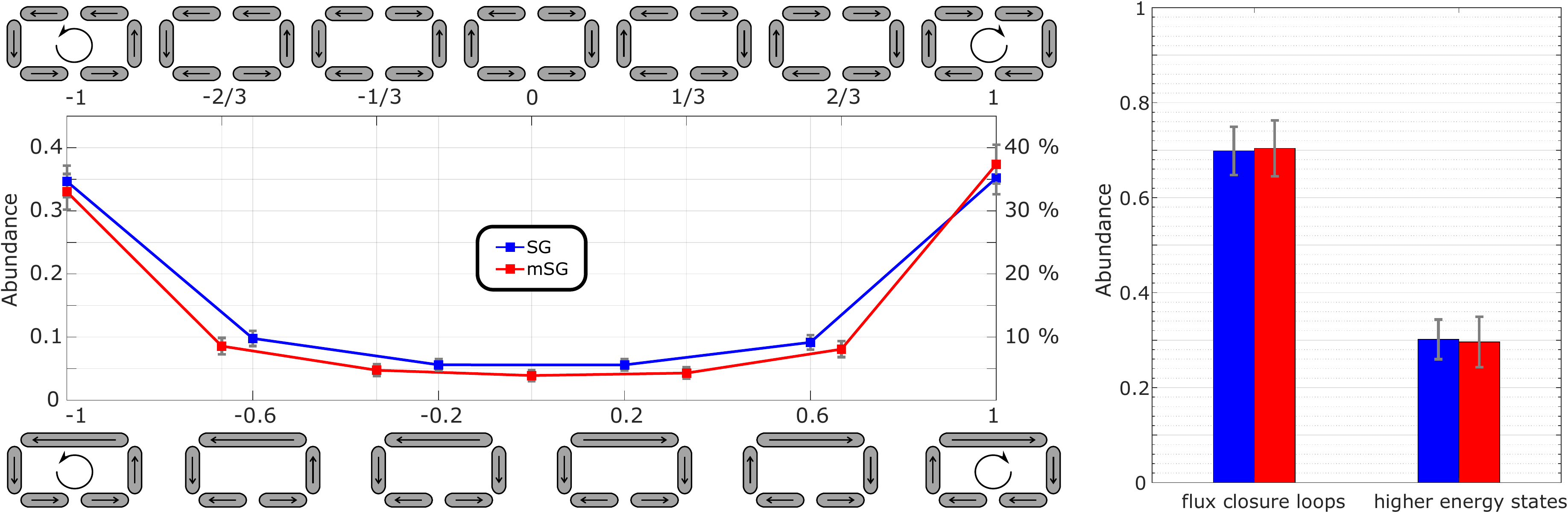}
\caption{{\bf Mesospin circulation in the SG and mSG lattice.} The circulation for loops of 5-mesospin in the SG lattice (blue), compared to the circulation for 6-mesospin loops in the mSG lattice (red) are shown in the left graph. The circulation of the loops was normalized to +1 (-1) for a total clockwise (anti-clockwise) alignment of all mesospins of a loop. This normalization results in a change of the loop circulation by $\pm$0.4 for each mesospin flipped in the SG lattice and by $\pm\frac{1}{3}$ in the mSG lattice. The right graph illustrates the degeneracy corrected abundance of flux closure loops and higher energy states for both the SG and mSG.}
\label{fig:chirality}
\end{figure*}

\subsection{Vertex abundances}

Statistical analysis of the vertex configurations of the mesospins allows for some insight into the role of the interactions and the degree of order in our arrays. The vertex-types, illustrated on the right side of Fig.~\ref{fig:Stats}, have different degeneracies:  type-I$_{4}$ has a degeneracy of 2, while type-II$_{4}$, type-III$_{4}$ and type-IV$_{4}$ have a degeneracy of 4, 8 and 2 respectively. To illustrate the difference from a random distribution we show the degeneracy corrected abundance for the fourfold and threefold coordinated vertices. To this end, we divide the vertex counts by their degeneracy, normalizing these new degeneracy-corrected vertex counts to one. The degeneracies for the threefold coordinated vertices are 2, 4 and 2 for type-I$_{3}$, type-II$_{3}$ and type-III$_{3}$, respectively. The larger number of low energy vertex configurations as seen in Fig.~\ref{fig:Stats} clearly shows that the mesospins are interacting in both SG and mSG lattices.
However, the number of higher excitations in both the fourfold and threefold vertices are a manifestation of thermal activity during the freezing process. These high energy vertex states (type-II$_4$-IV$_4$ and type-II$_3$-III$_3$, see Fig. \ref{fig:Stats} for vertex type illustration) have been observed earlier using the same magnetic material. \cite{ShaktiME} Whilst the distribution of vertex states are not random, it is still not possible to identify significant differences between the SG and mSG lattices. Therefore, we conclude that on the length-scale of the fourfold and threefold coordinated vertices no significant difference between the SG and mSG lattice can be identified, along with no measurable influence of the different energy-scales on the magnetic ordering. So far we have investigated the correlations of the magnetisation direction of neighbouring mesospins as well as the correlations of three and four interacting mesospins at a vertex. The next length-scale at which the anisotropy impacts the emergent behaviour is, therefore,  across several interacting mesospins and vertices.

\subsection{Flux closure loops}

\begin{figure*}[t!]
\centering
\includegraphics[width=2\columnwidth]{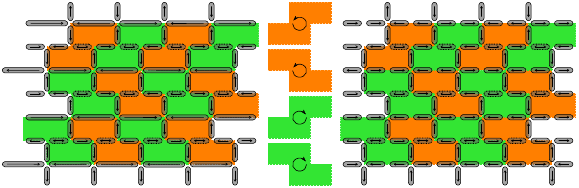}
\caption{{\bf Proposed ground state configuration in the SG (left) and mSG lattice (right)}. This ground state configuration takes into account only the vertex energies, while ignoring the different activation energies for short- and long-mesospins. All fourfold-coordinated vertices are in their lowest energy state. This configuration leads to a frustration among the threefold-coordinated vertices, as 50\% of them are in their lowest energy state but the other 50\% are in their first excited state, equivalent to the Shakti lattice \cite{Chern2013PRL}. The proposed ground state creates flux closure loops, as unit-cells of the ground state manifold. The four different flux closure loops are coloured orange or green, depending on their sense of flux circulation (sense of rotation). The border mesospins, 8~in the SG and 10~in the mSG lattice, are head to tail aligned, with an either clockwise or anti-clockwise sense of rotation. The non-border mesospin is frustrated and located in the center of the flux closure loop. The magnetic orientation of this island has no impact on the overall energy of the entire lattice, but it defines the position of the excited threefold-coordinated vertex.}
\label{fig:groundstate}
\end{figure*}

Flux closure loops are a way to minimize the energy of the lattice by reducing the residual stray fields. The flux closure loops in the SG and mSG lattice are defined as follows: The smallest flux closure loop in the SG lattice consists of one long- and four short-mesospins and in the mSG  of six short-mesospins, as depicted graphically in Fig.~\ref{fig:chirality}. The lowest energy fully flux closed loops are defined as having a circulation $+1$ for a clockwise and $-1$ for an anti-clockwise flux closure.  
Energetically, the most favorable circulations will be the fully closed loops with a circulation of $\pm1$. For higher energy states, where one or more islands are reversed, the normalised circulation is modified by $\pm$0.4 for each mesospin flipped in the SG lattice and by $\pm\frac{1}{3}$ in the mSG lattice. In Fig.~\ref{fig:chirality}, we compare the abundance of full and partial flux closure for the SG and the mSG lattices. 
The low energy flux closure loops, with values of $\pm1$ have the same abundance in the SG and the mSG lattice. The higher energy states, with one or more reversed mesospins at first glance appear more abundant in the SG lattice, but combining all flux closure loops (circulation values $\pm1$) to one value and summing up all remaining higher energy states to another value, it becomes clear that there are no significant differences between the flux closure loops in the SG and the mSG lattice (right graph in Fig \ref{fig:chirality}). Judging from this loop circulation evaluation in Fig.~\ref{fig:chirality} and the vertex statistics in Fig. \ref{fig:Stats}, we can assume that the SG and mSG lattice are similarly ordered and there is no major differences between them, originating from the different energy-scales. Thus far, we have only observed ordering and correlation at the short-range level, such as the vertex configurations and the smallest possible flux closure loops. To identify the next length-scale, we now turn our attention to the collective ground state of the SG lattice.

\subsection{Ground state manifold} 

The analysis presented above, focusing at the short-range scale, clearly indicates that the energy-scales of the two sizes overlap and drive an elaborate ordering mechanism. We can  therefore, simplify the evaluation of the ground state manifold by only minimizing the vertex energies, not taking into account the two different activation energies for the long- and short-mesospins. This assumption results in the ground state manifold which we present in Fig.~\ref{fig:groundstate} for one of the possible ground state configurations for the SG and the mSG lattice.

All fourfold coordinated vertices in the SG ground state manifold are in their energetically lowest state, the type-I$_4$ state (see Fig.~\ref{fig:Stats} for state notations). The geometrical configuration of the SG lattice leads to a vertex-frustration at the threefold coordinated vertices, which is present at all temperatures. This vertex-frustration results in 50\% of all threefold coordinated vertices being in their lowest energy state (type-I$_3$), with the remaining 50\% in their first excited state (type-II$_3$). The SG lattice geometry also yield a high degree of degeneracy, as the arrangement of the ground state unit-cells (illustrated in Fig.~\ref{fig:groundstate}) can be achieved in numerous ways. In addition, the central mesospin, inside these ground state unit-cells, defines the position of the type-II$_3$ excitation and is energetically degenerate. In a real sample, in which the dynamics of the freezing process itself also play a role, excitations and localised disorder are inevitably locked into the system which prevents long-range ground states from forming. This has been observed in the vertex statistics presented for these  (Fig.~\ref{fig:Stats}) and similar artificial arrays such as the Shakti lattice \cite{ShaktiME}.

\begin{figure*}[t!]
\centering
\includegraphics[width=2\columnwidth]{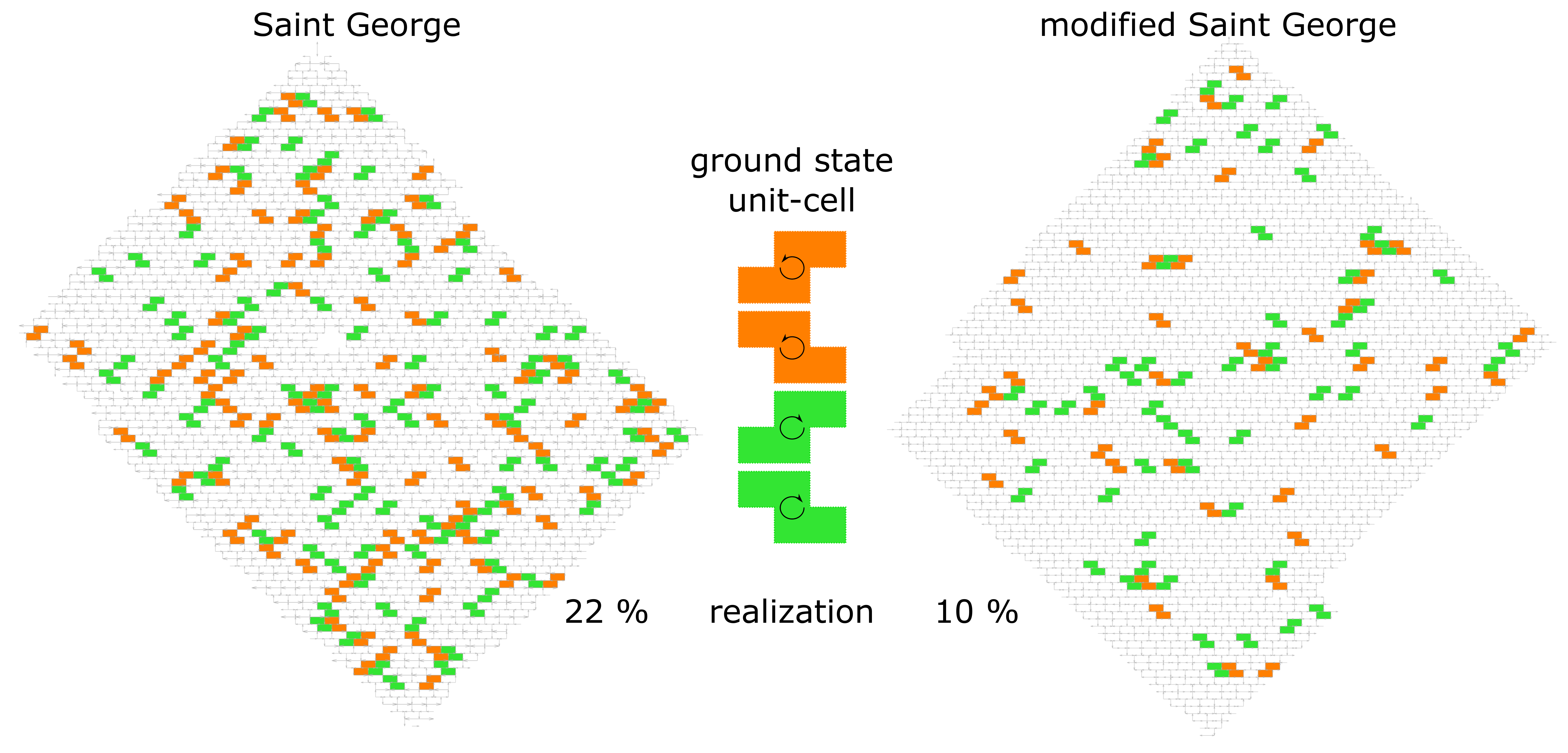}
\caption{{\bf Mapping of ground state unit-cells in the SG and mSG lattice.} The ground state unit-cells introduced in Fig.~\ref{fig:groundstate} are mapped out experimentally in our arrays. There is a clear difference in the quantity of ground state unit-cells found in the SG and the mSG lattice. A 22\% ground state realization is observed in the the SG lattice, with a 10\% realization in the mSG lattice.}
\label{fig:GSmaps}
\end{figure*}

All of the analyses until now took place in the short-range and amongst two, three, four, five or six mesospins, residing in the short-range order. Looking in more detail at the ground state unit-cell, it becomes evident that flux closure loops consisting of two long- and six short-mesospins for the SG lattice and ten short-mesospins for the mSG lattice should also be considered when discussing the overall ordering of the lattice geometry. Furthermore, these ground state unit-cells with their 8 or 10 mesospins represent an intermediate length-scale within the SG lattice. As the ground state unit-cells can be arranged in multiple ways throughout the lattice geometry, the flux closure loops can not easily be investigated in the same fashion as we did with the circulation values in Fig.~\ref{fig:chirality}. It is possible however, to investigate and compare directly their observed abundances in the lattices.

\subsection{Magnetic ordering on an intermediate length-scale}

In pursuit of this magnetic ordering on an intermediate length-scale, we turn our attention towards the ground state unit-cells in our measured arrays (Fig.~\ref{fig:PEEM}). In this way, we are able to detect the first major differences between the SG and the mSG lattices, by counting the number of ground state unit-cells present in the arrays. Fig.~\ref{fig:GSmaps} illustrates the distribution of these ground state unit-cells in the measured lattices and in this representation the fact that more ground state unit-cells can be found in the SG lattice, than in the mSG lattice, is clearly highlighted. Evaluating further these ground state unit-cells quantitatively, we find 22\% ground state realization for the SG lattice but only 10\% ground state realization for the mSG lattice. This difference is attributed to the interplay of the two energy-scales, where the long-mesopins influence the ordering of the short-mesospins on length-scales beyond that of nearest neighbours. The highly susceptible short-mesospin matrix acts as an interaction enhancer and propagates the influence of the long-mesospins into the intermediate length-scale, which is reflected in the result presented in Fig. \ref{fig:GSmaps} and related ground state abundances. Accounting for the amount of elements present in a ground state unit-cell one can argue that the retrieved values are just a reflection of the degrees of freedom, which are reduced in the SG lattice when compared with the mSG lattice.

\subsection{Conditional mesospin arrangement}

\begin{figure}[b!]
\centering
\includegraphics[width=1\columnwidth]{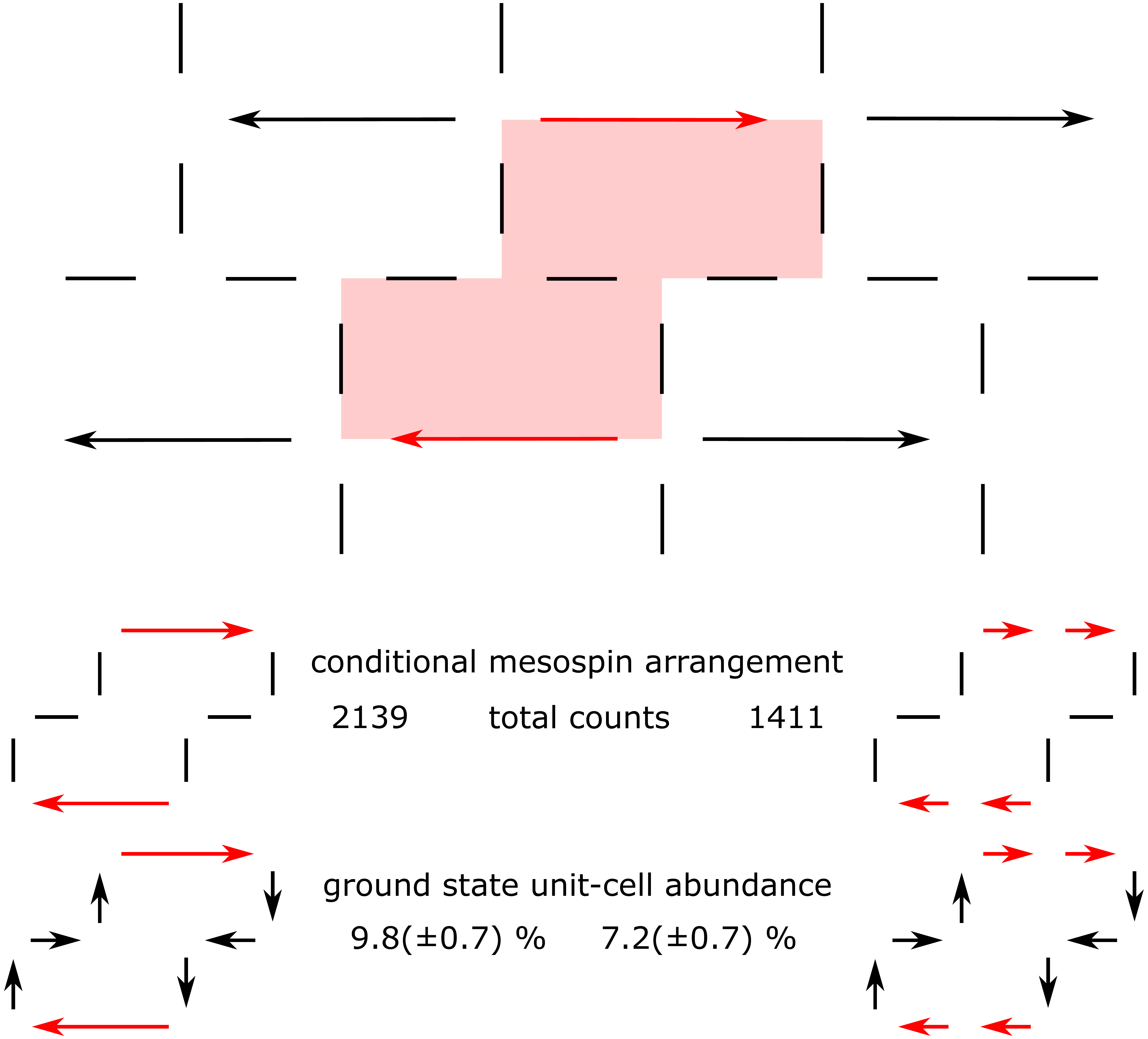}
\caption{{\bf Conditional arrangement of the ground state unit-cells in the SG and mSG lattice.} The two long-mesospins from subsequent long-mesospin chains are anti-ferromagnetically aligned to each other. In this conditional mesospin arrangement we compare the abundance of the ground state unit-cells in the SG and mSG lattice and see a clear evidence of a stronger ordering in the SG lattice.}
\label{fig:ConSA}
\end{figure}

We have also investigated a conditional arrangement of mesospins wherein the two short-mesospins representing the long-mesospin in the mSG lattice have to be ferromagnetically aligned to each other (see Fig. \ref{fig:ConSA}). As a first step we identify two long-mesospins in two subsequent long-mesospin chains in the SG lattice (red mesospins in Fig. \ref{fig:ConSA}). These mesospins must have an anti-parallel alignment of their magnetization direction ignoring the orientation of the six short-mesospins connecting the two. In order, to have the same pre-defined condition in the mSG lattice we perform the same selection, but this time we additionally choose that the two short-mesospins (corresponding to the long-mesospin in the SG lattice) to be ferromagnetically aligned to each other while the magnetization direction in the subsequent chains is still anti-parallel oriented. The magnetization of the long-mesospins (or two short-mesospins in the mSG) in the subsequent mesospin chains is set in the conditional arrangement and we focus our attention to the subset of the six short-mesospins connecting the selected mesospin configuration. In this way we restrict the degrees of freedom in this subset of mesospins in the mSG to be the same as that in the SG lattice. Searching for arrangements where two long-mesospins of subsequent long island chains are anti-ferromagnetically aligned to each other, we observed a total of 2139 and 1411 instances in the SG and mSG lattices respectively. The different abundance for this conditional arrangement in the SG and mSG demonstrating the influence of the degrees of freedom. By investigating these conditional arrangements for the ground state unit-cells we are comparing a subset of mesospins (six short-mesospins), which is identical in the SG and mSG lattice. If these six mesospins are randomly arranged we would receive an abundance of  $\frac{1}{2^6} = 1.6 \%$. Investigating the conditional arrangements for ground state unit-cells an abundance of $9.8 (\pm0.7) \%$ for the SG and and $7.2 (\pm0.7) \%$ for the mSG can be found. These values are clearly distinguishable from a random arrangement, considering their uncertainty and can therefore be attributed to the interaction energies in these lattices. It is striking to see a difference between the SG and mSG lattice only at this length-scale. The different abundance of ground state unit-cells in the conditional mesospin arrangements can only be explained by the difference in activation energy for the short- and long-mesospins. The conditional mesospin arrangement shows therefore the direct influence of this activation energy on the intermediate-range ordering in mesoscopic spin systems.

\section{Conclusion}

The analysis of the results of this study, indicate that the long-mesospins act as ordering seeds, around which the short-mesospins preferably align themselves, which is in agreement with previous studies of the Shakti lattice \cite{ShaktiME}. In contrast to the latter case though, here the short-mesospins affect the character of the magnetic order amongst the long-mesospins, through an apparently stronger interaction between the two distinct mesospin energy-scales, arising from the lattice geometry. These fluctuating Ising-like short-mesospins can be seen as interaction modifiers for the long-mesospin chains, in a similar way as recently presented by \citet{Ostman_NatPhys_2018}, where the placement of magnetic discs in the center of the square artificial spin ice vertex, alters the strength and the degeneracy of the magnetic coupling between the Ising-like mesospins. In the case of the SG lattice, the short-mesospins are placed in between the long-mesospins but with a horizontal offset. During the cooling process the long-mesospins strongly interact with the highly susceptible fluctuating short-mesospin matrix, modifying the normally ferromagnetic correlations between the long-mesospins. As such, the interplay between the energy-hierarchy and the topology needs to be considered as the resulting magnetic order can not be understood by simply following a strict separation between the short- and long-mesospin ordering. A domination of the energy-hierarchy over the topological influence would favour predominantly the formation of ferromagnetic chains along the long-mesospin chains, which we do not observe in these samples. A detailed investigation of the exact ordering mechanism during the freezing process in such systems, as for example while varying the cooling rate, would shed more light on the interplay between the different energy-scales.

Focusing on the short-range order, we observed that there are no significant differences between the SG and the mSG lattice, neither at the vertex abundances nor the smallest flux closure loops length-scale. These observations hint that the two energy-scales have no impact on the magnetic ordering, but turning our attention to the intermediate length-scale we see a significant lower ground state realization in the mSG lattice, which is independent of the degrees of freedom. We therefore conclude that the two step freezing process in the SG lattice allows the mesospin system to achieve twice as much ground state realization as in the one step freezing system of the mSG lattice. This can be explained as an intermediate-range ordering originating from the long-mesospins, while the short-range order is dominated by the short-mesospins. The intermediate length-scale order is improved in the SG lattice by the long-mesospins interacting with the short-mesospin matrix during the freezing process, which is finally expressed in the higher degree of order on this length-scale. While the short-range order is purely dominated by the activation energy of the short-mesospins, reflected in non-distinguishable magnetic ordering in the SG and mSG at the short-range.

Our study highlights the importance of the energy-scales in combination with the topology on the collective magnetic order in artificial ferroic systems. A full understanding of this interplay between energy-scales and topology, makes it possible to use artificial magnetic nanostructures with differently sized \cite{ShaktiME} and shaped elements \cite{Arnalds2014APL, Ostman_NatPhys_2018}, in order to design systems where emergence and frustration can be studied in a systematic way at the mesoscale \cite{Anderson:1972dn, Nisoli_Kapaklis_NatPhys2017}. Furthermore, this work calls for the development and studies of appropriate models accounting for the presence of multiple energy-scales \cite{Wilson:1979wn}, in settings where geometry and/or topology have a significance.  This approach may have also long-term importance for the design of arrays with enhanced magnetic reconfigurability, which can be utilized for instance in magnonics \cite{Gliga_PRL2013, Bhat:2018dj, Ciuciulkaite:2019km, Gypens_2018}, exploiting the dependence of their dynamic magnetization spectra on their micromagnetic states \cite{Jungfleisch:2016fa, Lendinez_review_2020, Sloetjes_arXiv_2020}.

\begin{acknowledgements}
The authors acknowledge support from the Knut and Alice Wallenberg Foundation project `{\it Harnessing light and spins through plasmons at the nanoscale}' (2015.0060), the Swedish Foundation for International Cooperation in Research and Higher Education (Project No. KO2016-6889) and the the Swedish Research Council (Project No. 2019-03581). The patterning was performed at the Center for Functional Nanomaterials, Brookhaven National Laboratory, supported by the U.S. Department of Energy, Office of Basic Energy Sciences, under Contract No. DE-SC0012704. This research used resources of the Advanced Light Source, which is a DOE Office of Science User Facility under contract No. DE-AC02-05CH11231. U.B.A. acknowledges funding from the Icelandic Research Fund grants Nr. 207111 and 152483 and the University of Iceland Research Fund.
The authors would also like to express their gratitude to Dr. Erik \"Ostman for assistance with PEEM-XMCD
measurements and valuable conversations, Dr. David Greving for help with the electron-beam lithography, as well as to Dr. Cristiano Nisoli (Los Alamos National Laboratory, U.S.A.) and Dr. Ioan-Augustin Chioar (Yale University, U.S.A.) for fruitful discussions and valuable feedback. The excellent on-site support by Dr. Andreas Scholl and Dr. Rajesh V. Chopdekar,  at the 11.0.1 PEEM3 beamline of the Advanced LightSource, in Berkeley, California, USA, is also greatly acknowledged.
\end{acknowledgements}


%

\end{document}